\documentclass[12pt,fleqn,twoside]{article}
\usepackage{espcrc1,epsfig}

\usepackage{graphicx}
\usepackage[figuresright]{rotating}

\newcommand{\AmS}{{\protect\the\textfont2
  A\kern-.1667em\lower.5ex\hbox{M}\kern-.125emS}}
\def\nn{\nonumber}
\def\als{\alpha_s}
\def\be{\begin{equation}}
\def\ee{\end{equation}}
\def\bea{\begin{eqnarray}}
\def\eea{\end{eqnarray}}

\def\gev{\,{\rm GeV}}
\newcommand{\aem}{\alpha_{\rm em}}

\newcommand{\LQCD}{\Lambda_{\rm{QCD}}}
\newcommand{\da}{{distribution amplitude}}
\newcommand{\das}{{distribution amplitudes}}
\newcommand{\wf}{wave function}
\newcommand{\wfs}{wave functions}
\newcommand{\spa}{soft physics approach}
\renewcommand{\d}{\rm d}
\newcommand{\ibid}[1]{{\it ibid.}~#1}

\newcommand{\lsim}{\raisebox{-3pt}{$\,\stackrel{\textstyle <}{\sim}\,$}}
\newcommand{\gsim}{\raisebox{-3pt}{$\,\stackrel{\textstyle >}{\sim}\,$}}

\title{PROBING THE NUCLEON AT LARGE MOMENTUM TRANSFER}

\author{P. Kroll\address{Fachbereich Physik, Universit\"at Wuppertal\\ 
        Gau\ss strasse 20, D-42097 Wuppertal, Germany}%
        \thanks{ Supported in part by the TMR network ERB 4061 Pl 95 0115
                }}

\begin{document}

\sloppy
\pagestyle{empty}

\begin{flushright}
WU B 99-18 \\
hep-ph/9907301\\[20mm]
\end{flushright}

\begin{center}
{\Large\bf PROBING THE NUCLEON AT LARGE MOMENTUM TRANSFER }\\[20mm]

{\Large P.\ Kroll}\\[10mm]
{\it Fachbereich Physik, Universit\"at Wuppertal}\\[1mm]
{\it Gau\ss strasse 20, D-42097 Wuppertal, Germany}\\[1mm]%
{ E-mail: kroll@theorie.physik.uni-wuppertal.de}\\[20mm]
\end{center}

\begin{center}
Invited talk presented at the NUCLEON99 workshop\\[5mm] 
Frascati (June 1999)
\end{center}
\newpage

\setcounter{page}{1}

\maketitle

\begin{abstract}
The central role of soft nucleon matrix elements in reactions of
high energy electrons or real photons with nucleons  is
emphasized. These soft matrix elements are described in terms of skewed
parton distributions. Their connections to ordinary parton
distributions, form factors, Compton scattering and hard
meson electroproduction is discussed.  
\end{abstract}

\section{INTRODUCTION}

\begin{figure}[t]
\parbox{\textwidth}{\begin{center}
   \psfig{file=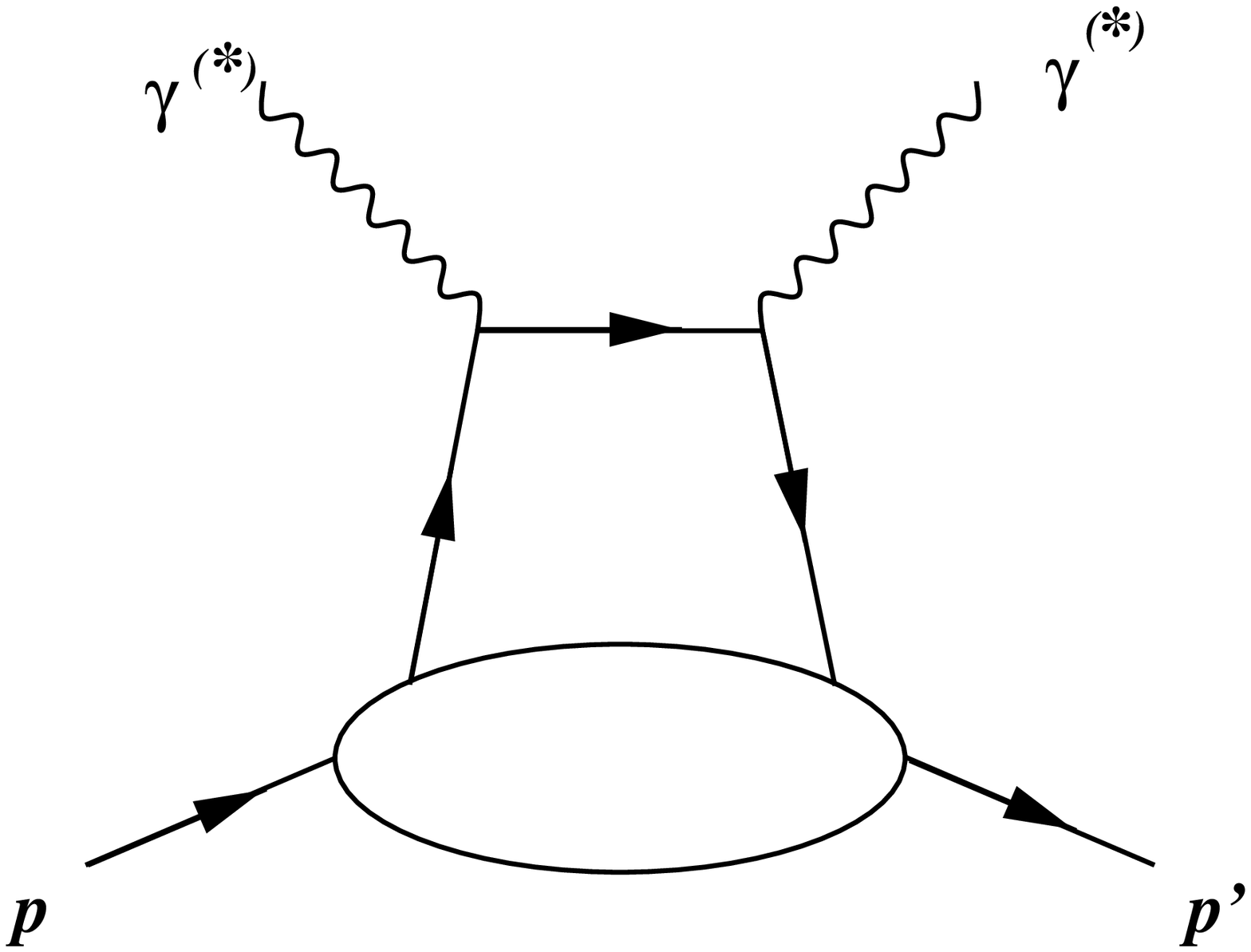,%
          bbllx=45pt,bblly=230pt,bburx=545pt,bbury=610pt,%
           width=4.0cm,clip=} \hspace{2cm}
   \psfig{file=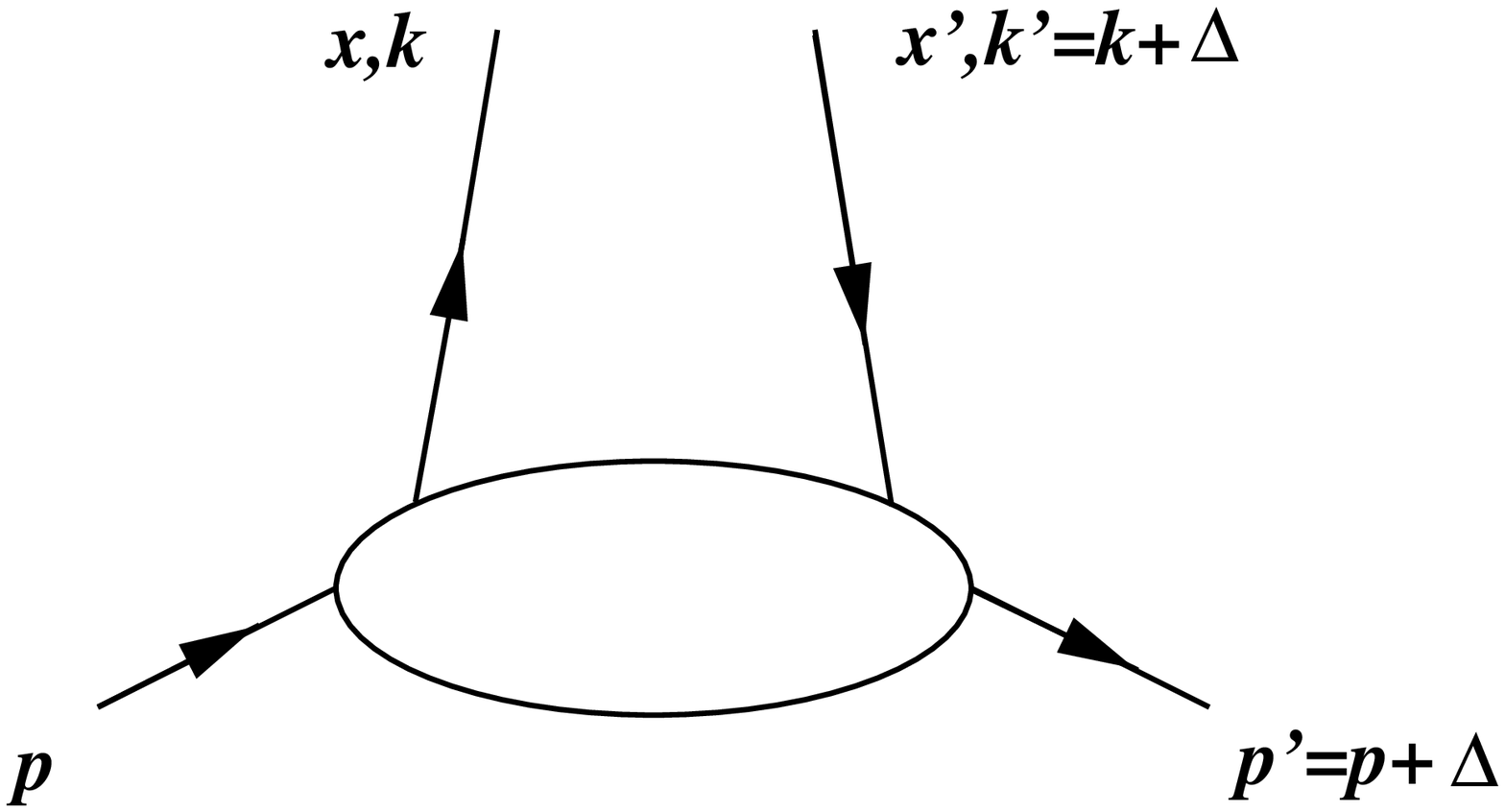,%
          bbllx=30pt,bblly=275pt,bburx=565pt,bbury=565pt,%
           width=4.5cm,clip=}
\vspace{-1.0cm}
\end{center}}
\caption{The handbag diagram (left) and the kinematics for SPDs (right).}
\vspace*{-0.5cm}
\label{fig:handbag}
\end{figure}

Interactions of the nucleon with high energy electromagnetic probes, either
electrons (i.e.\ virtual photons) or  real photons, are described
by the handbag diagram shown in Fig.~\ref{fig:handbag}. Depending on
the virtuality of the incoming photon, $Q^2$, and on the momentum
transfer from the incoming to the outgoing nucleon, $t$, different
processes are described by the handbag diagram: For forward 
scattering, $t=0$, with highly virtual photons, $Q^2\gg m^2$, (
$m$ being the mass of the nucleon) the diagram represents the total cross
section for the absorption of virtual photons by nucleons. This is
the domain of deep inelastic lepton-nucleon scattering (DIS) from which we learn
about the ordinary unpolarized, $q^a(x)$, and polarized, $\Delta
q^a(x)$, parton distribution functions (PDF). If one considers 
outgoing real photons the handbag describes deeply virtual Compton
scattering (DVCS) for $t\simeq 0$, $Q^2\gg m^2$ as well as real (RCS) or
virtual (VCS) Compton scattering for $-t \gg m^2$ (and $-u \gg m^2$)
and $Q^2$ either zero or non-zero, respectively.
Common to all these processes are soft nucleon matrix elements which
are parameterised in terms of skewed parton distributions (SPDs),
i.e.\ generalized PDFs. The same soft nucleon matrix elements are also 
relevant to the electromagnetic form factors of the nucleon 
and to hard meson electroproduction at large $Q^2$ and small $-t$. The
handbag diagram and in particular the soft nucleon matrix elements are of
major interest here at the NUCLEON99 workshop as the program
reveals. A synopsis of the physics related to the soft nucleon matrix
elements will be attempted in this talk.

\section{SPDs}

The SPDs attracted the interest of theoreticians only
recently. Despite of this a large number of papers already appeared
in which properties and applications of SPDs are discussed.

Defining the kinematics as in Fig.\ \ref{fig:handbag} and introducing the
fractions of light-cone-plus momentum components 
\be
x=k^+/p^+, \qquad\qquad x'=k'{}^+/p'{}^+, \qquad\qquad \zeta=-\Delta^+/p^+,
\label{fra}
\ee
one may define SPDs as ($t=\Delta^2$) \cite{mue98,ji97,rad97}
\be
p^+ \int {{\d} z^-\over 2\pi}\, e^{i\, x p^+ z^-}
          \langle p'|
               \overline\psi{}_{a}(0)\, \gamma^+\,\psi_{a}(z^-)  
                                          | p\rangle \,=\,
 {\cal F}^a_\zeta(x;t)\, \bar{u}(p') \gamma^+ u(p)
                   + {\cal K}^a_\zeta\, \bar{u}(p')
                        \frac{i\sigma^{+}_{\rho}\Delta^\rho}{2m} u(p).\!\! 
\label{spd}
\ee
$\psi_a$ denote the field operator of a flavour-$a$ quark. 
The decomposition (\ref{spd}) provides two SPDs, ${\cal
F}^a_\zeta$ and ${\cal K}^a_\zeta$, in analogy to the Dirac, $F_1$, and
Pauli, $F_2$, form factors of the nucleon. The corresponding matrix element of
$\gamma^+\gamma_5$ defines two further SPDs, ${\cal G}^a_\zeta$ and
${\cal L}^a_\zeta$. Note that the three momentum fractions defined in
(\ref{fra}) are not independent of each other:
\be
       x'\,=\, (x-\zeta)/(1-\zeta) \,.
\ee
Hence, the SPDs are functions of only three variables, e.g.\ $x$,
$\zeta$ and $t$. For
$\zeta\leq x \leq 1$ they describe the emission of a quark with
momentum fraction $x$ from the nucleon and the absorption of a quark
with momentum fraction $x'$. In the region $0\leq x <\zeta$ the
nucleon with momentum $p$ emits a quark-antiquark pair and is left as
a nucleon with momentum $p'=p+\Delta$. Re-interpreting a quark with
negative momentum fraction as an antiquark with positive momentum
fraction one finds that the region $-1+\zeta\leq x \leq 0$ describes
the emission and absorption of antiquarks. The symmetry relation
\be
    {\cal F}^{\bar{a}}_\zeta (x;t) = - {\cal F}^a_\zeta (\zeta -x;t)
\ee
holds as well as similar ones for the other SPDs. One may also consider helicity or
flavour non-diagonal SPDs. In \cite{fel99}, for instance, $b-u$ SPDs
have been discussed which control $B \to\pi$ transitions. 

The definition (\ref{spd}) reveals the close relationship of SPDs to
ordinary PDFs and to electromagnetic form factors. For instance,
\be
{\cal F}^a_{\zeta=0} (x;t=0)\,=\, q^a(x)\,,  \quad
{\cal G}^a_{\zeta=0} (x;t=0)\,=\, \Delta q^a(x)\,, \quad
\int_{-1+\zeta}^1 {\d}x\, {\cal F}^a_{\zeta} (x;t) \,=\, F_1^a (t)\,.
\label{red-spd-q}
\ee
The zeroth order moments of the other three SPDs, ${\cal K}^a_{\zeta}$,
${\cal G}^a_{\zeta}$ and ${\cal L}^a_{\zeta}$, provide
the contributions of flavour-$a$ quarks to the Pauli, 
axial and pseudoscalar form factors,
respectively. The full form factors are given by appropriate sums, for instance,
\be
    F_1(t)\,=\, \sum_a e_a\, F_1^a(t)\,.
\ee
The SPDs ${\cal K}^a_\zeta$ and ${\cal L}^a_\zeta$ 
which are not accessible in DIS,
provide new information on the nucleon, e.g.\ about 
the orbital angular momentum the quarks carry.

From the reduction formulas (\ref{red-spd-q}) it
is obvious that the SPDs establish a link between inclusive and
exclusive reactions. As the PDFs the SPDs are universal, i.e.\
process-independent functions. Proofs of factorisation into soft and
hard parts have been given
for DVCS in Refs.\ \cite{rad97,ji98} and for hard meson
electroproduction (large $Q^2$, small $-t$) in
Ref.\ \cite{CFS}. The latter process is dominated by longitudinally
polarized virtual photons for $Q^2\to \infty$; the cross section for
transversely polarized photons is suppressed by $1/Q^2$.     

\begin{figure}[t]
\parbox{\textwidth}{\begin{center}          
   \psfig{file=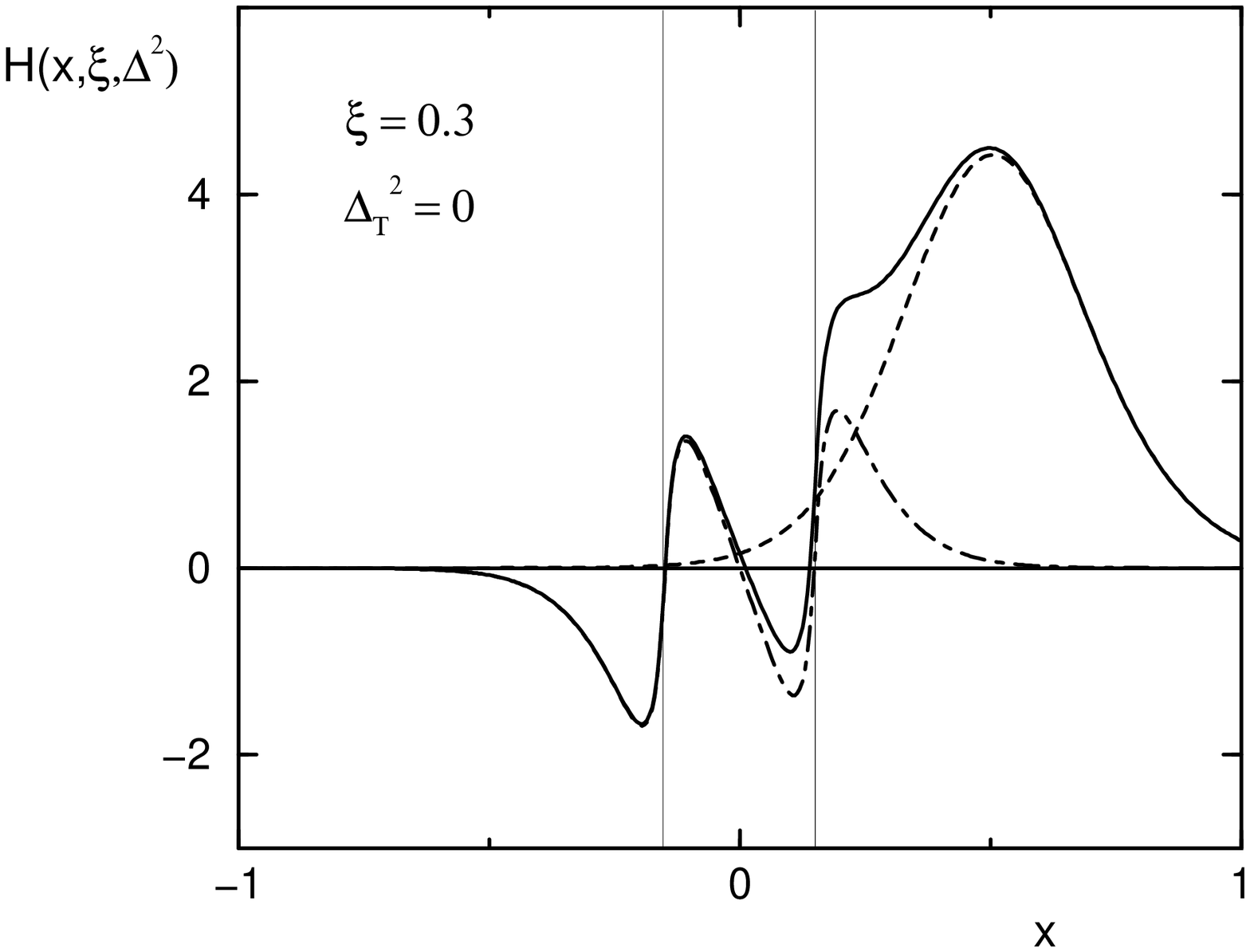,%
          bbllx=30pt,bblly=45pt,bburx=560pt,bbury=445pt,%
           width=5.5cm,clip=}\hspace{1.9cm}
    \psfig{file=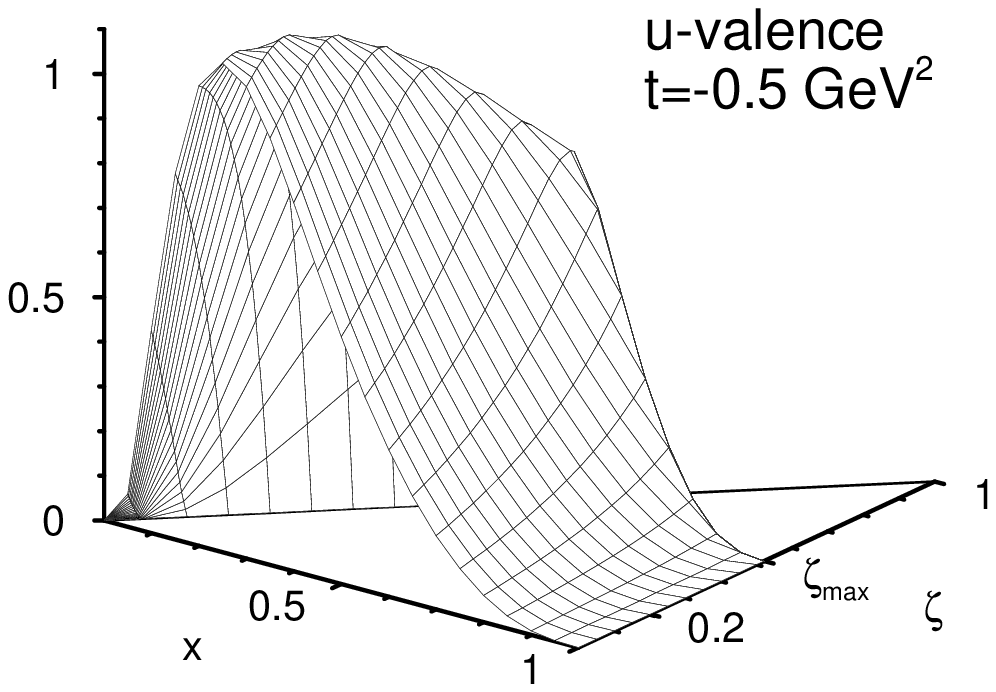, width=5.5cm}
\vspace{-1cm}
\end{center}}
\caption{Left: Predictions of the instanton model \cite{pet98} for the $u+d$
      SPD $H(x_{ji},\xi,\Delta^2)$ in Ji's notation \cite{ji97}
    ($x_{ji}=(2x-\zeta)/(2-\zeta)$; $\xi=\zeta/(2-\zeta)$). Dashed
    (dash-dotted, solid) line: valence (sea, total) contributions.
    $\Delta_T^2 \equiv -\Delta^2-\xi^2 m^2 = 0$. Right: Predictions of
    the soft physics approach \cite{DFJK} for the $u$-valence SPD ${\cal
    F}^u_\zeta -{\cal F}^{\bar{u}}_\zeta $ in the range $\zeta\leq x\leq 1$.} 
\label{fig:bochum}
\vspace*{-0.5cm}
\end{figure}
\begin{figure}
\parbox{\textwidth}{\begin{center}
     \psfig{file=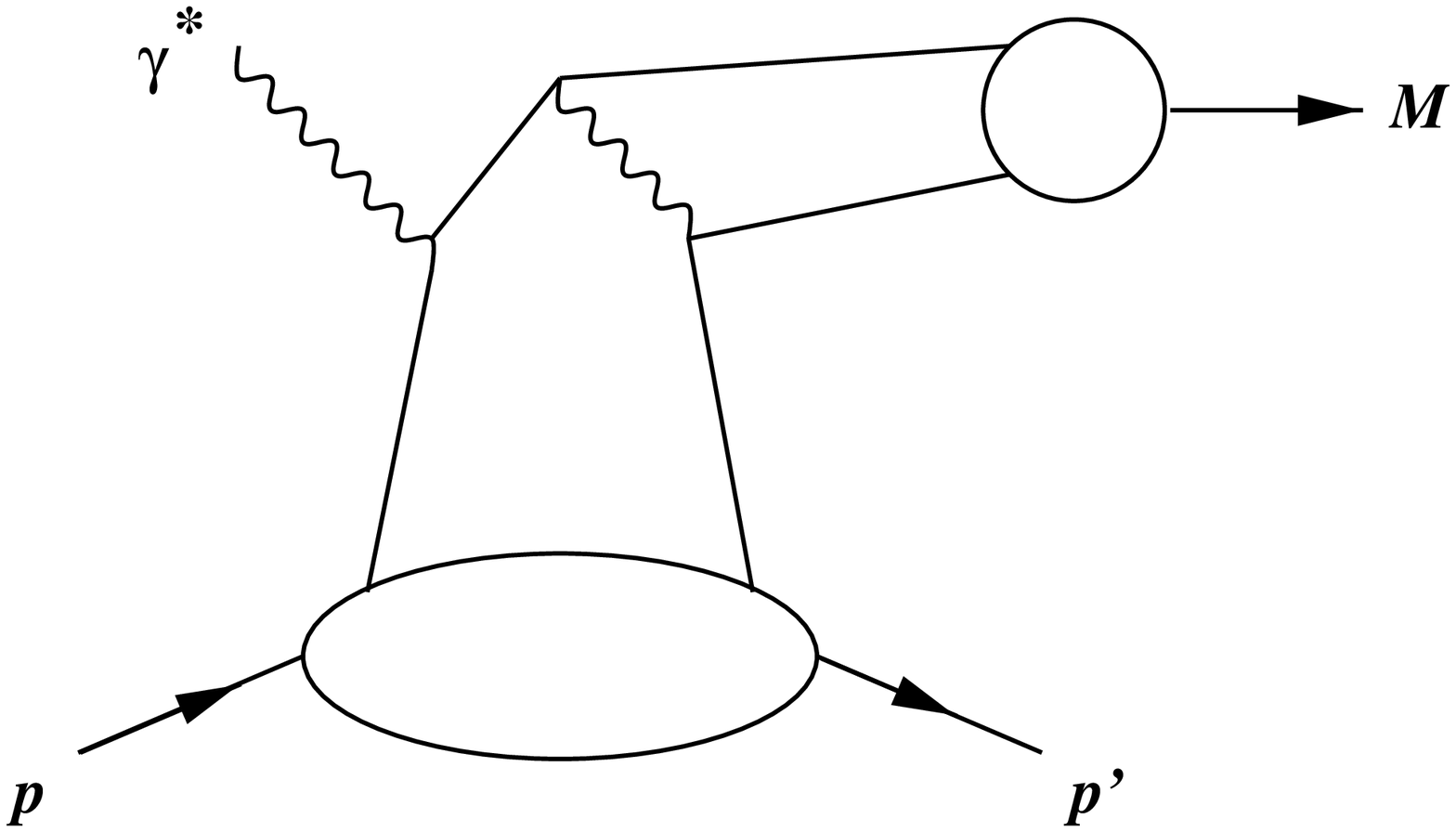, 
       bbllx=15pt,bblly=265pt,bburx=550pt,bbury=580pt,width=4.5cm,clip=}\hspace{2cm}
   \psfig{file=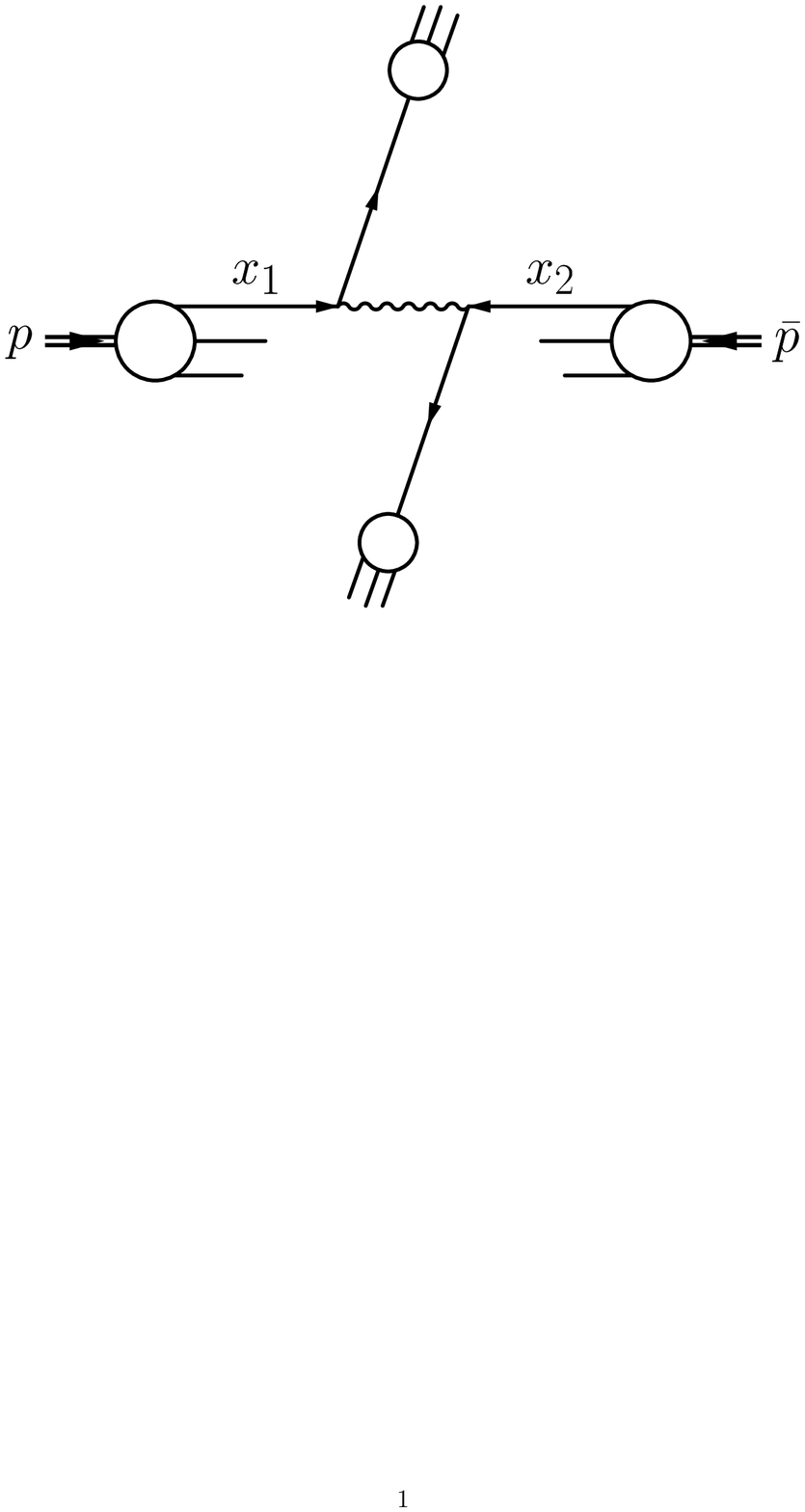,%
          bbllx=105pt,bblly=470pt,bburx=490pt,bbury=765pt,%
           width=4.5cm,clip=}
\vspace{-1cm}
\end{center}}
\caption{Left: A typical Feynman diagram for hard electroproduction of
   mesons. Right: Lowest order jet production in $p\bar{p}$ collisions.}
\label{fig:spd1}
\vspace*{-0.5cm}
\end{figure}

As the usual PDFs they SPDs should be extracted from data. However,
in the present situation of complete absence of any DVCS data and
only scarce information on hard electroproduction of mesons this is
not possible and we therefore have to rely on models. As examples
results on SPDs from the instanton model \cite{pet98} and from the
soft physics approach proposed in \cite{DFJK} are shown in Fig.\
\ref{fig:bochum}. Occasionally,
the SPDs are parameterized as products of form factors,
PDFs and, perhaps, an additional rational function of $x$ and
$\zeta$. On the basis of such parameterizations predictions
for DVCS \cite{van99} and for $\gamma_L^* p\to M p$, e.g.\
\cite{van99,man98}, have been worked out and compared to the few
existing data on the latter process. Diffractive $J/\Psi$
electroproduction at small $x$-Bjorken is a particular interesting
case because this process is controlled by the gluon SPD \cite{rad96}.
A Feynman diagram for hard meson electroproduction is shown in Fig.\
\ref{fig:spd1}. The amplitude for vector meson production is
proportional to
\bea
{\cal M}^V &\sim& \sum_a \left[ \int {\d}z \frac{\Phi_V(z)}{z}\right]\; 
                                  \int_0^1 {\d}x\,  
                   \left\{ {\cal F}_\zeta^a(x;t) + 
                        {\cal F}_\zeta^{\bar{a}}(x;t) \right\}
                         \left\{ \frac{1}{x-i\epsilon} 
                         + \frac{1}{x-\zeta+i\epsilon}\right\} \nn\\
     &+&\; {\cal K}-{\rm terms}\,, 
\label{mes-pro}
\eea
where $\Phi_V$ is the usual vector meson distribution amplitude
characteristic of hard exclusive reaction. The amplitude for the
production of pseudoscalar mesons is obtained from (\ref{mes-pro}) by
replacing ${\cal F}^a_\zeta$ and ${\cal K}^a_\zeta$ through ${\cal
G}^a_\zeta$ and ${\cal L}^a_\zeta$. Thus, the complementary study of
hard vector and pseudoscalar meson electroproduction allows to
distantangle the two sets of SPDs from each other. To DVCS, on the
other hand, all four SPDs contribute.

\section{PARTON DISTRIBUTIONS}

Clearly, the SPDs are a field that is still in its infancies and where
theory is well ahead of experiment. This situation is to be contrasted
with that of the ordinary PDFs where detailed analyses of DIS, the Drell-Yan
process and $W^{\pm}$ production in proton-antiproton collisions have
been performed with the help of the DGLAP equation actively for many
years \cite{GRV,MRST,CTEQ}. An enormous progress in the quality of the
phenomenological PDFs has been achieved in recent years although still
a number of problems has to be settled up (e.g.\ the large- and
small-$x$ behaviour, the difference between the ${u}$ and
$d$ sea quark distributions, polarized PDFs). At any rate what remains
to be understood are the input PDFs ($g$, $u_v$, $d_v$, $\bar{u}$ and
$\bar{d}$) at some small scale $\mu_0$. Lattice QCD provides a few
moments of the PDFs with admittedly large uncertainties
\cite{goe96}. Results for PDFs at a low scale have been obtained from the
instanton model \cite{dia96} and from a constituent quark model
\cite{sco98}.

In order to demonstrate the present quality of the available PDFs, one may
compare the D0 \cite{abo98} and the preliminary CDF data
\cite{cdf} for the inclusive jet cross section in
$p\overline{p}$ collisions at $\sqrt{s}=1.8\, {\rm TeV}$ with
predictions from pQCD obtained through (cf.\ Fig.\ \ref{fig:spd1})
\be
{\d}\sigma^{p\overline{p}\to {\rm jet} X}\,=\,
     \frac{1}{\pi}\,\sum_{abc}\,\int\,{\d}x_1 {\d}x_2\, 
    q^a(x_1,\mu_F)\, q^b(x_2,\mu_F)\, {\d}\hat{\sigma}^{ab\to cX} \,,
\ee
and evaluated from present PDFs \cite{MRST,CTEQ}. ${\d}\hat{\sigma}$ is
the partonic cross section calculated to order $\als^3$ and $\mu_F$ a
factorisation scale of the order of the transverse jet energy,
$E_\perp$. 
Quark-quark scattering dominates the jet cross section at large
$E_\perp$ while, for moderately large $E_\perp$, substantial
contributions come from quark-gluon scattering. Although the general
agreement between experiment and theory is very good, a closer
inspection reveals discrepancies of the order of $10-20\%$; the CDF
collaboration, in particular, observes an excess of the jet rate at
the largest measured $E_\perp$. The reason for these discrepancies
is, perhaps, an insufficient accuracy of the valence and gluon PDFs at
$x\gsim 0.6$. In this regard it is important to realize that in most
of the analyses \cite{GRV,MRST,CTEQ} a vanishing $d/u$ ratio is
assumed in the limit $x\to 1$. However, as pointed out in
\cite{tom97}, correcting the deuteron data for nuclear binding effects
one obtains a $d/u$ ratio of about 0.2 from a re-analysis of
DIS. Finite values of the $d/u$ ratio have also been found in several
models \cite{DFJK,sim98}. An enhanced $d$-quark distribution at 
large $x$ may lead to a larger jet cross section at high $E_\perp$.

An improved understanding of the large-$x$ behaviour of the PDFs is
mandatory for searches of new physics. 

\section{EXCLUSIVE REACTIONS WITHIN pQCD}

\begin{figure}
\parbox{\textwidth}{\begin{center}
   \psfig{file=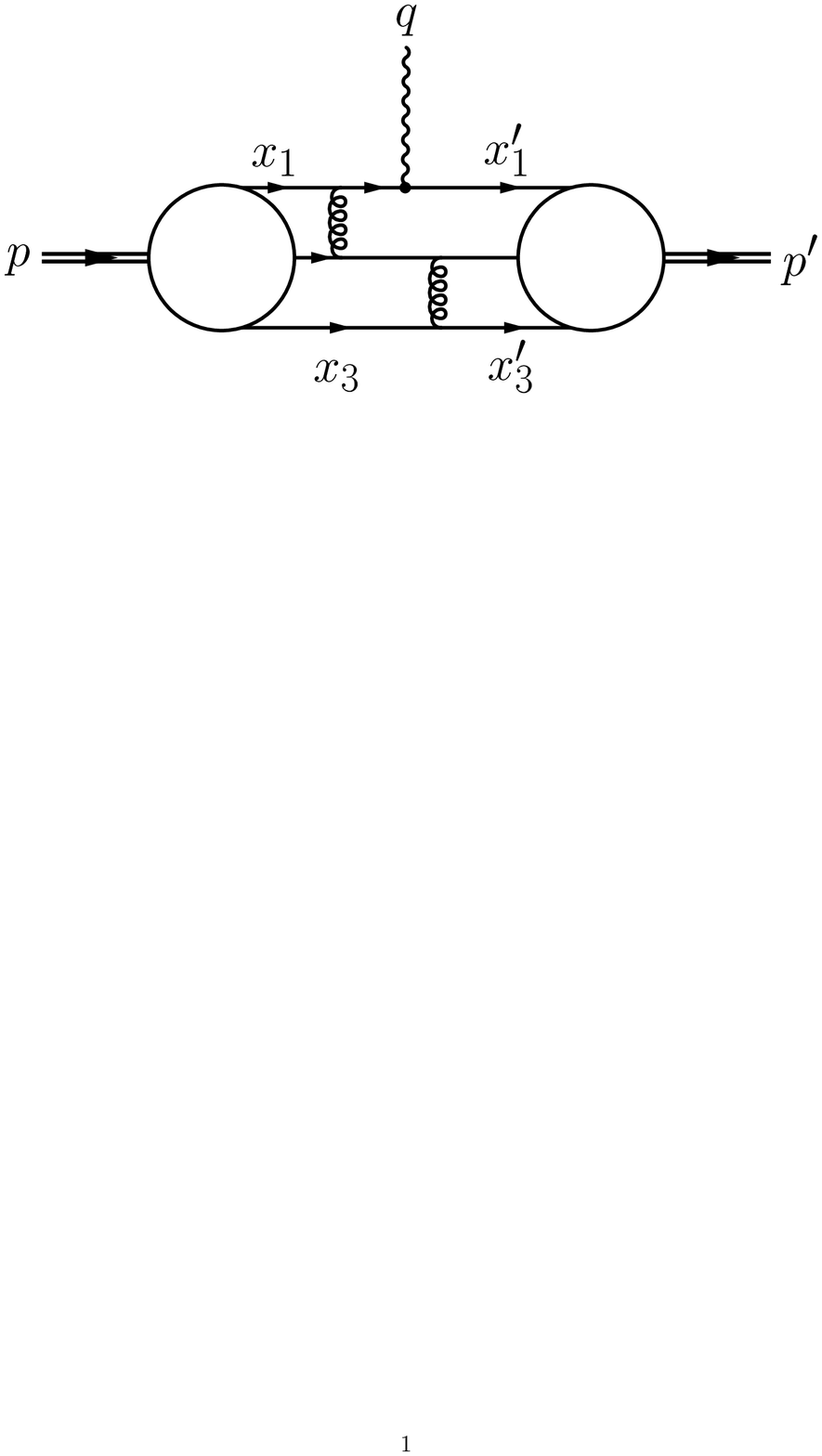,%
          bbllx=90pt,bblly=580pt,bburx=510pt,bbury=790pt,%
           width=5.0cm,clip=}\hspace{1.5cm}
   \psfig{file=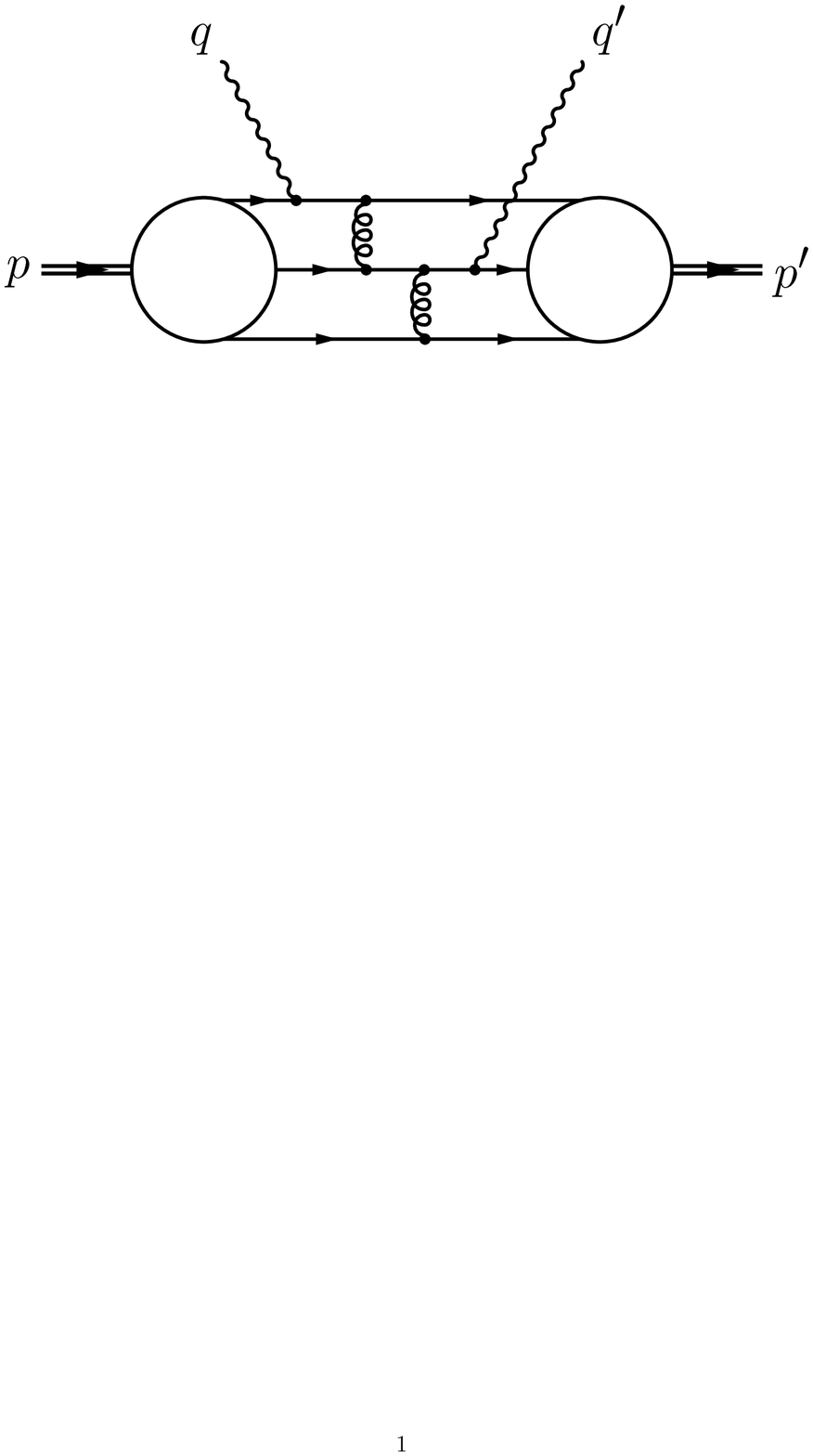,%
          bbllx=90pt,bblly=585pt,bburx=510pt,bbury=790pt,%
           width=5.0cm,clip=}
\vspace*{-1.0cm}
\end{center}}
\caption{Sample Feyman graphs for form factors (left) and Compton
scattering (right)  within pQCD.}
\vspace*{-0.5cm}
\label{fig:pqcd}
\end{figure}

Let me now turn to nucleon form factors, RCS and VCS at large momentum
transfer. QCD provides three contributions from the valence Fock state to
these processes, namely a soft overlap term with an active quark and two
spectators, the asymptotically dominant perturbative contribution
where by means of the exchange of two hard gluons the quarks are kept
collinear with respect to their parent nucleons (see Fig.\
\ref{fig:pqcd}) and a third contribution that is intermediate between
the soft and the perturbative contribution where only one hard gluon
is exchanged and one of the three quarks acts as a spectator. Both the
soft and the intermediate terms represent power corrections to the
perturbative contribution. Higher Fock state contributions are
suppressed. The crucial question is now what are the relative
strengths of the three contributions at experimentally accessible
values of momentum transfer, i.e.\ at $-t$ of the order of 10 GeV$^2$?
The pQCD followers assume the dominance of the perturbative
contribution and neglect the other two contributions while the soft
physics community presumes the dominance of the overlap
contribution. Which group is right is not yet fully decided although
comparison with the pion case \cite{kro96} seems to favour a strong
overlap contribution.

In pQCD \cite{bro80} the nucleon form factors and the Compton
amplitudes (see Fig.\ \ref{fig:pqcd}) are described by convolutions of
hard scattering amplitudes, to be calculated within pQCD to a given
order of $\als$, and distribution amplitudes $\Phi$ which represent
the nucleon's valence Fock state \wf{} integrated over transverse
momenta up to a factorisation scale $\mu_F$ of order $-t$:
\be
   F_1 ({\cal M}) \sim f_N^2(\mu_F)\sum_\beta \int [{\d}x]_3 [{\d}x']_3\,
                         \Phi_\beta(x',\mu_F) \,
                      T_{H\beta}^{F({\cal M})}(x,x',t)\, \Phi_\beta(x,\mu_F).
\label{eq:pqcd}
\ee
$\beta$ labels different spin-flavour combinations of the
quarks in the three-particle Fock state. $[{\d}x]_3$ is the
three-particle integration measure.

Considering only \wfs{} with zero orbital angular momentum component in
the direction of the nucleon's momentum, one can show that there is
only one independent scalar \wf{} or \da{}, say 
$\Phi_{123}$ for the $u_+ u_- d_+$-configuration in a proton with
positive helicity. The \das{} for other parton configurations are to
be obtained from $\Phi_{123}$ by appropriate permutations of the
indices. The \da{} is subject to evolution and turns into $120 x_1 x_2
x_3$ for an asymptotically large scale $\mu_F$.

An immediate consequence of the pQCD approach to hard exclusive
reactions are the power laws (i.e.\ the dimensional counting rule
behaviour):
\be
   F_1 \sim t^{-2} (\ln{-t})^{-\gamma}, \hspace{3cm}
  \frac{{\d}\sigma}{{\d}t}(\theta \;{\rm fixed}) 
                                  \sim s^{-6} (\ln{\;s})^{-2\gamma}\,    
\label{eq:powers}
\ee
where $\theta$ is the cm scattering angle. $\gamma$ is fed by $\als$
as well as by the evolution of both $f_N$ and the dis\-tri\-bution
amplitude. The latter contribution, while it is clearly positive, cannot be 
specified in general. Thus, $\gamma \geq 2 [1 + (2/(3\beta_0)] \simeq 2.05$.
While the powers quoted in (\ref{eq:powers}) are in fair agreement with
experiment even for surprisingly small values of momentum transfer or energy
there is no evidence for the logarithms. This is, by the way, true for a
large class of reactions. Occasionally it is argued that the effective
scales in these processes are so small that the running coupling
becomes frozen, i.e.\ the Landau pole is effectively cut off. This
indicates, however, that one is not in the perturbative regime. With a
frozen $\als$ the experimentally observed approximate power law
behaviour appears as a transient phenomenon that only 
holds in a limited range of momentum transfer, the perturbative logarithms
will become visible at very large momentum transfer. 

In the formal limit $\ln{(-t/\LQCD^2)}\to \infty$, in which the \da{}
evolves into its asymptotic form, pQCD predicts $F_1^p/F_1^n \to 0$
and a very small positive value for the neutron form factor. 
This is not at all what we see in the data. As a way out Chernyak and
Zhitnitsky \cite{CZ} assigned the difference between experiment and
asymptotic pQCD prediction to the form of the \da{}.
Indeed, for form factors \cite{CZ} and RCS 
\cite{niz91,van97} agreement with experiment is accomplished if
\das{} are utilized that are strongly concentrated in the end-point regions
where one of the momentum fractions, $x_i$, tends to zero. 
The drawback of this solution is that the bulk of the
perturbative contribution is accumulated in regions where the
virtualities of the internal gluons are well below 1
GeV$^2$, i.e.\ in regions where pQCD is not applicable. 
A second drawback of the end-point region concentrated \das{} is that
they, if merged with, say, Gaussian transverse momentum dependences in 
light-cone \wfs{}, lead to large overlap contributions
\cite{isg,bol96} which exceed the data by huge factors.

The first drawback can be cured if the modified perturbative approach
is exploited. In this approach that has been developped by Botts, Li
and Sterman \cite{bot89},   
the transverse momentum dependence of the hard scattering
amplitude and of the wave functions is retained and Sudakov
suppressions are taken into account. 
Applications to the form factor revealed that this calculation
is self-consistent in the sense that the bulk of the perturbative
contribution is now accumulated in regions of sufficiently small $\als$
\cite{ber95,jai99}. However, the suppression of the end-point regions
is so strong that the perturbative contribution to the nucleon form
factor is much smaller than experiment even for the end-point region
concentrated \das{}
\cite{ber95}. In Ref.\ \cite{jai99} a larger perturbative contribution
is obtained by neglecting the intrinsic transverse momentum
dependence of the \wf{} and allowing for a more complicated infrared
cut off than in \cite{ber95}. Nevertheless
the question remains to be answered what to do with the large
soft contributions?

\begin{figure} 
\parbox{\textwidth}{\begin{center}
       \psfig{file=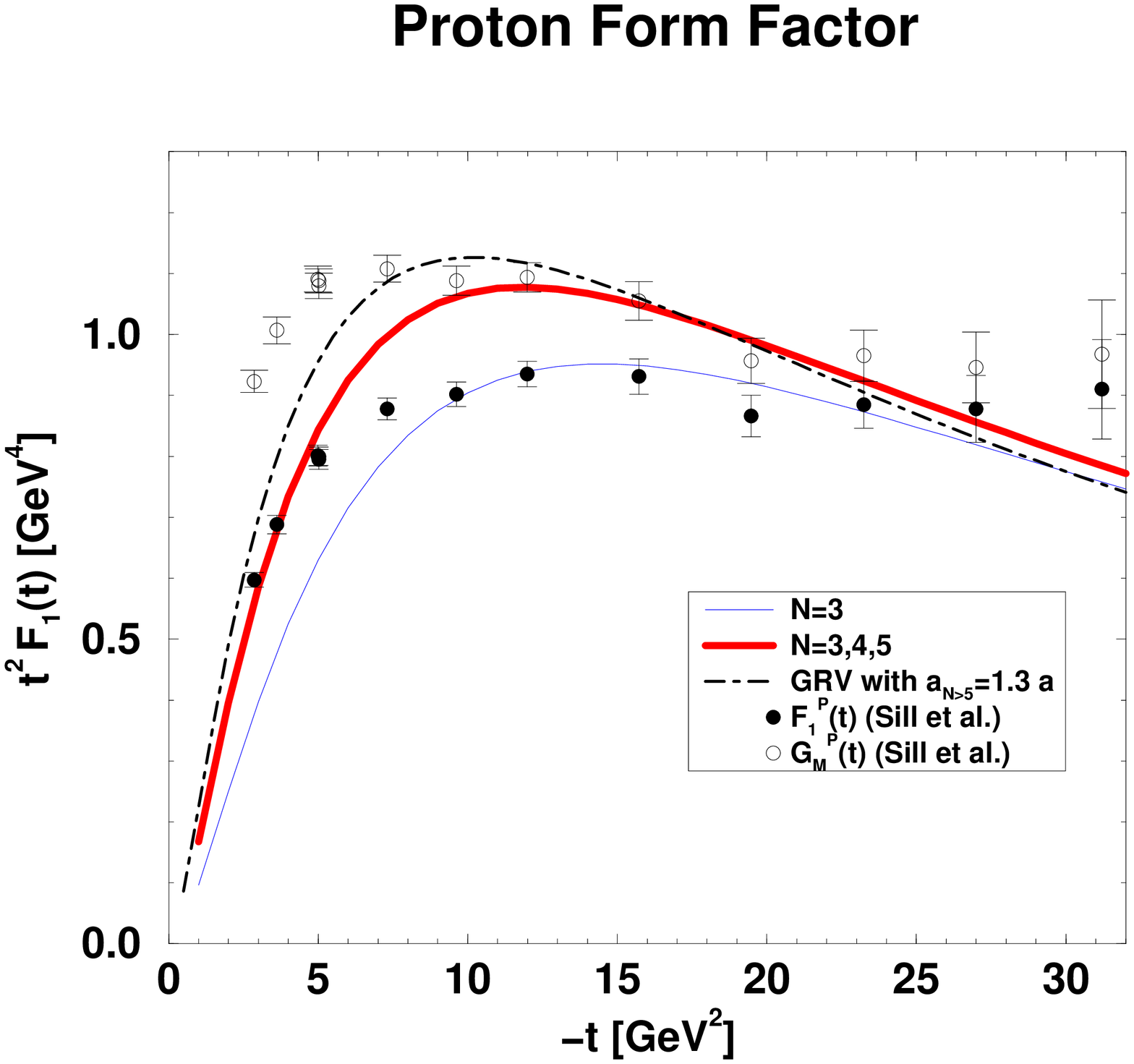, 
          bbllx=100pt,bblly=0pt,bburx=590pt,bbury=635pt,%
           width=5.5cm, angle=-90, clip=} \hspace{0.5cm}
       \psfig{file=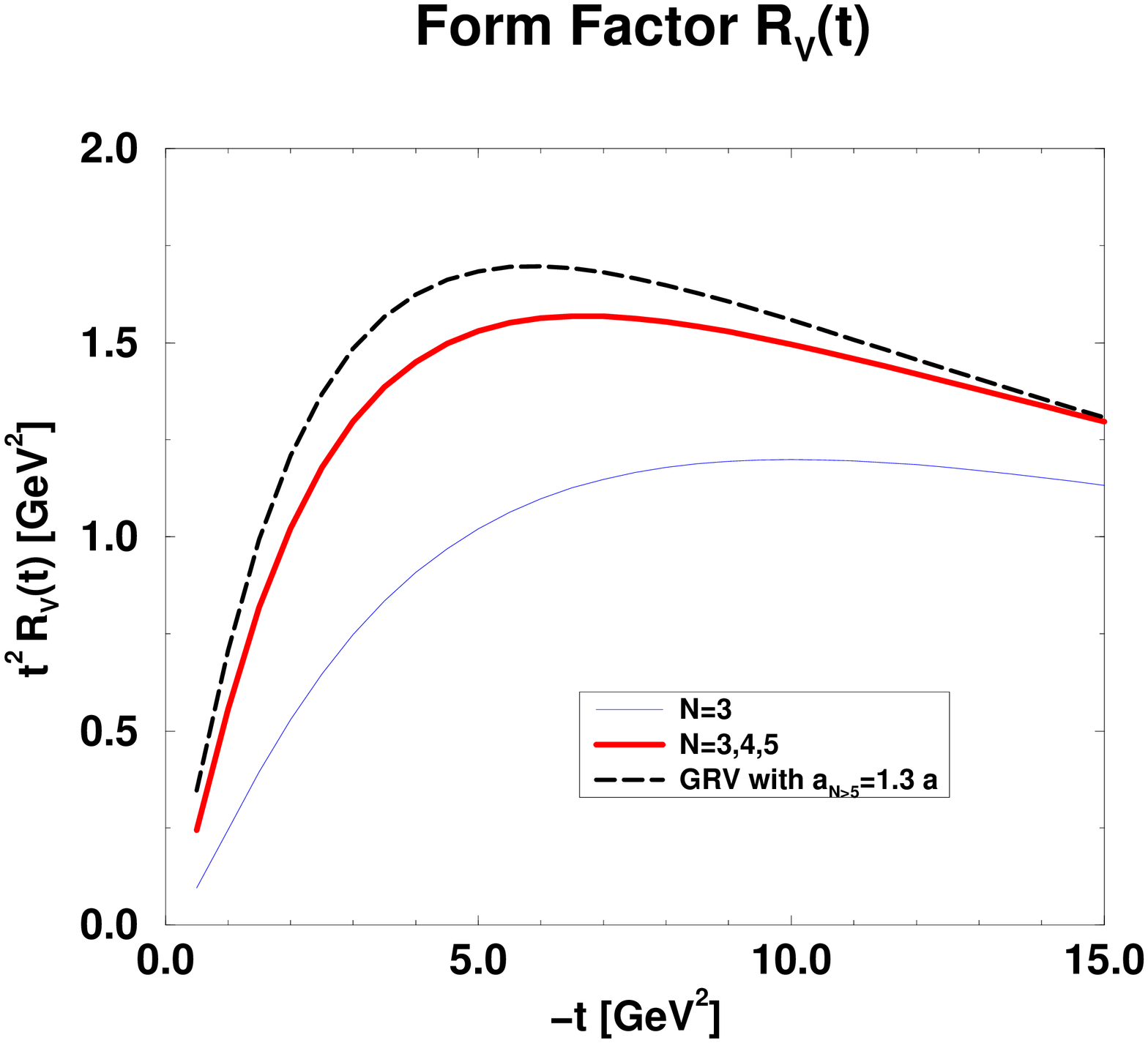, 
          bbllx=90pt,bblly=0pt,bburx=590pt,bbury=655pt,%
           width=5.5cm, angle=-90, clip=} 
\vspace*{-1cm}
\end{center}}
\caption{The Dirac (left) and the vector Compton (right) form factors
   of the proton as predicted by the soft physics approach
   \cite{DFJK,bol96}. Data are taken from \cite{sil93}. The data on
   the magnetic form factor, $G_M$, are shown in order to demonstrate
   the size of spin-flip effects.} 
\vspace*{-0.5cm}
\label{fig:soft}
\end{figure}

A new \da{} for the nucleon has been proposed in \cite{bol96}:
\be 
 \Phi^{BK}_{123} \,=\, 60\, x_1x_2x_3\, (1+3x_1)
\label{eq:BK}
\ee
valid at a factorisation scale of 1 GeV. Combining it with a Gaussian
transverse momentum dependence, one finds soft contributions in fair
agreement with the data \cite{sil93} on nucleon form factors, see Fig.\
\ref{fig:soft}. On the other hand, the perturbative contribution
evaluated from this \da{} within the modified perturbative approach is
as small as only a few $\%$ of the experimental value. Therefore, this
is to be regarded as a consistent calculation in which the soft
contribution clearly dominates for experimentally accessible values of
momentum transfer. 

\section{THE SOFT PHYSICS AP\-PROACH TO COMPTON SCATTERING}

For $s, -t, -u\gg m^2$ the handbag diagram shown in Fig.\
\ref{fig:handbag} describes RCS and VCS. To see this it is of
advantage to choose a frame of reference in which $\Delta^+=0$
(implying $\zeta=0$ and $t=-\Delta^2_\perp$). In order to evaluate the
SPD appearing in the handbag diagram one may use a Fock state
decomposition of the nucleon and sum over all possible spectator
configurations. The crucial assumption is then that the soft hadron
\wfs{} are dominated by virtualities in the range $|k_i^2|\lsim
\Lambda^2$, where $\Lambda$ is a hadronic scale of the order of 1 GeV,
and by intrinsic transverse parton momenta, $k_{\perp i}$, defined
with respect to their parent hadron's momentum, that satisfy $k_{\perp
i}^2/x_i\lsim \Lambda^2$. Under this assumption factorisation of the
Compton amplitude in a hard photon-parton amplitude and $1/x$-moments
of SPDs can be shown to hold \cite{DFJK,rad98a}. 

As a consequence of this result the Compton amplitudes conserving the
nucleon helicity are given by
\be
{\cal M}_{\mu'+,\,\mu +} \,=\, \;2\pi\aem \left[{\cal
    H}_{\mu'+,\,\mu+}\,(R_V + R_A) \,
    + \, {\cal H}_{\mu'-,\,\mu-}\,(R_V - R_A) \right ]\,.
\label{final}
\ee
Nucleon helicity flip is neglected. $\mu$ ($\nu$) and $\mu'$ ($\nu'$) are
the helicities of the incoming and outgoing photon (nucleon) in the
photon-nucleon cms. The photon-quark subprocess amplitudes, ${\cal
H}$, are calculated for massless quarks to lowest order QED.
The soft nuclon matrix elements in Eq.\ (\ref{final}), $R_V$ and
$R_A$, represent form factors specific to Compton scattering
\cite{DFJK,rad98a}. $R_V$ is defined by
\begin{eqnarray} 
\lefteqn{\sum_a e_a^2\, \int_0^1\, \frac{{\d} x}{x}\, p^+
   \int {{\d} z^-\over 2\pi}\, e^{i\, x p^+ z^-} 
           \langle p',\nu'|\,
     \overline\psi{}_{a}(0)\, \gamma^+\,\psi_{a}(z^-) - 
     \overline\psi{}_{a}(z^-)\, \gamma^+\,\psi_{a}(0) 
     \,| p,\nu\rangle}  \nn \\[0.5em]
&=& R_V(t)\, \bar{u}(p',\nu')\, \gamma^+ u(p,\nu)\,
 + R_T(t)\, \frac{i}{2m}\bar{u}(p',\nu')
                          \sigma^{+\rho}\Delta_\rho u(p,\nu)\,.
\label{R-form-factors}
\eea
$R_T$ being related to nucleon helicity flips, is neglected in
(\ref{final}). There is an analogous equation for the axial vector
nucleon matrix element, which defines the form factor $R_A$. 
Due to time reversal invariance the form factors $R_V$ and $R_A$ 
are real functions. As the definition (\ref{R-form-factors}) reveals
they are $1/x$ moments of SPDs at zero skewedness parameter
$\zeta$. As in DIS and DVCS, only the plus
components of the nucleon matrix elements enter in the Compton
amplitude, which is a nontrivial dynamical feature given that, in
contrast to DIS and DVCS, not only the plus components of the nucleon
momenta but also their minus and transverse components are large now. 

As shown in Ref.\ \cite{DFJK} the SPDs can be represented as
generalized Drell-Yan light-cone \wf{} overlaps in the region
$\zeta\le x \le 1$. Assuming a plausible Gaussian $k_{\perp 
i}$-dependence of the soft Fock state \wfs{}, one can explicitly carry
out the momentum integrations in the Drell-Yan formula. For simplicity
one may further assume a common transverse size parameter, $\hat a$,
for all Fock states. This immediately allows one to sum over them,
without specifying the $x_i$-dependence of the \wf{}s. One then
arrives at \cite{DFJK,rad98a} 
\bea
F_1(t)&=& \sum_a\, e_a\, \int {\d} x\, 
         \exp{\left[\frac12 \hat a^2 t \frac{1-x}{x}\right]}   
                      \{ q_a(x) - \bar{q}_a(x) \} \,, \nn\\[0.5em]
R_V(t)&=& \sum_a\, e_a^2\, \int \frac {{\d} x}{x}\, 
         \exp{\left[\frac12 \hat a^2 t \frac{1-x}{x}\right]} 
                          \{ q_a(x) + \bar{q}_a(x) \} \,,
\label{ffspd}
\eea
and the analogue for $R_A$ with $q_a+\bar{q}_a$ replaced by $\Delta
q_a + \Delta \bar{q}_a$. A similar representation is also obtained for
the nucleon's axial form factor. The form factor representation
(\ref{ffspd}) is very instructive as it elucidates the link between
the parton distributions of DIS and exclusive reactions. Note that the
combination $\sum e^2_a\: [q_a +\bar{q}_a]$ appearing in $R_V$ is just
the leading twist contribution to the structure function $F_2$ of DIS
divided by $x$, the corresponding term in $R_A$ corresponds to $2g_1$.

The only parameter appearing in (\ref{ffspd}) is the effective
transverse size parameter $\hat{a}$; it is known to be about 1
GeV{}$^{-1}$ with an uncertainty of about 20$\%$. Thus, this parameter
only allows some fine tuning of the results. Evaluating, for instance,
the form factors from the PDFs derived by Gl\"uck et al. (GRV)
\cite{GRV} with $\hat{a}=1\, GeV^{-1}$, one already finds good results
\cite{DFJK}. Improvements are obtained by treating the lowest three
Fock states explicitly with specified $x$-dependences (e.g. with the
\da{} (\ref{eq:BK})). Results for the $u$-valence SPD, 
$t^2 F_1$ and $t^2 R_V$ obtained that way in Ref.\ \cite{DFJK} are
displayed in Figs.\ \ref{fig:bochum}, and \ref{fig:soft},
respectively. Both the scaled form factors as well as $t^2 R_A$
exhibit broad maxima and, hence, mimic dimensional counting rule
behaviour in the $t$-range from about 5 to 15 GeV{}$^2$, set by the
transverse hadron size. The position, $t_0$, of the maximum of $t^2
F_i$, where $F_i$ is any of the soft form factors, is to be determined
from the implicit equation    
\begin{equation}
- t = 4 \hat a^{-2}\, \left\langle { \frac{1-x}{x}} \right\rangle^{-1}_{F_i,t}\,.
\label{maxpos}
\end{equation} 
The mean value $\langle \frac{1-x}{x} \rangle$ comes out
around $0.5$ at $t=t_0$. Hence, $-t_0 \simeq 8 \hat a^{-2} \simeq 8\,\gev^2$.
For very large momentum transfer the form factors turn gradually into
the soft physics asymptotics $\sim 1/t^4$. This is the region where the
perturbative contribution ($\sim 1/t^2$) takes the lead. 

The amplitude (\ref{final}) leads to the RCS cross section    
\be
\frac{{\d} \sigma}{{\d} t} = \frac{{\d} \hat{\sigma}}{{\d} t}
                       \left [\, \frac{1}{2} (R_V^2(t) + R_A^2(t))
        -\, \frac{us}{s^2+u^2}\, (R_V^2(t)-R_A^2(t)) \,\right] \,.
\ee
It is given by the Klein-Nishina cross section 
\be
      \frac{{\d} \hat{\sigma}}{{\d} t}\, = \,\frac{2\pi\aem^2}{s^2}\; 
                         \frac{s^2+u^2}{-us} \,,       
\ee
multiplied by a factor that describes the structure of the nucleon in
terms of two form factors. Evidently, if the form factors scale as
$1/t^2$, the Compton cross section would scale as $s^{-6}$ at fixed cm
scattering angle $\theta$. In view of the above discussion (see also
Fig.\ \ref{fig:soft}) one therefore infers that approximate
dimensional counting rule behaviour holds in a 
limited range of energy. The magnitude of the Compton cross section is
fairly well predicted as is revealed by comparison with the admittedly
old data \cite{shu79} measured at rather low values of $s$, $-t$ and
$-u$ (see Fig.\ \ref{fig:all}). Cross sections of 
similar magnitude have been obtained within the perturbative approach
\cite{niz91,van97} and within the diquark model \cite{kro91}. The
latter model is a variant of the standard perturbative approach in
which diquarks are considered as quasi-elementary constituents of the
nucleon. Better data are needed for a crucial test of
the \spa{} and its confrontation with other approaches.
\begin{figure}
\parbox{\textwidth}{\begin{center}
   \psfig{file=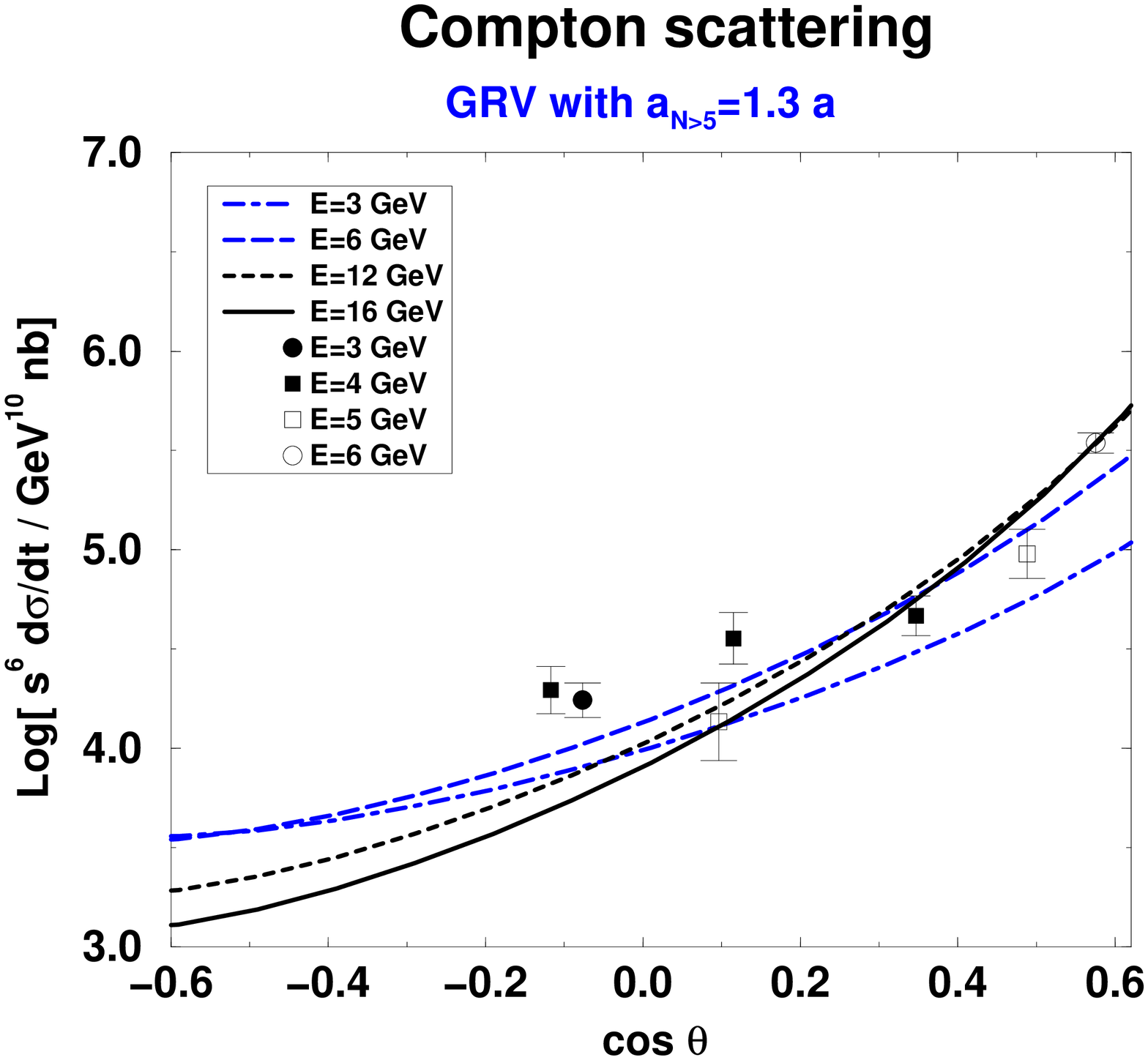, 
          bbllx=92pt,bblly=0pt,bburx=590pt,bbury=640pt,%
           width=5.7cm, angle=-90, clip=} \hspace{1cm}
   \psfig{file=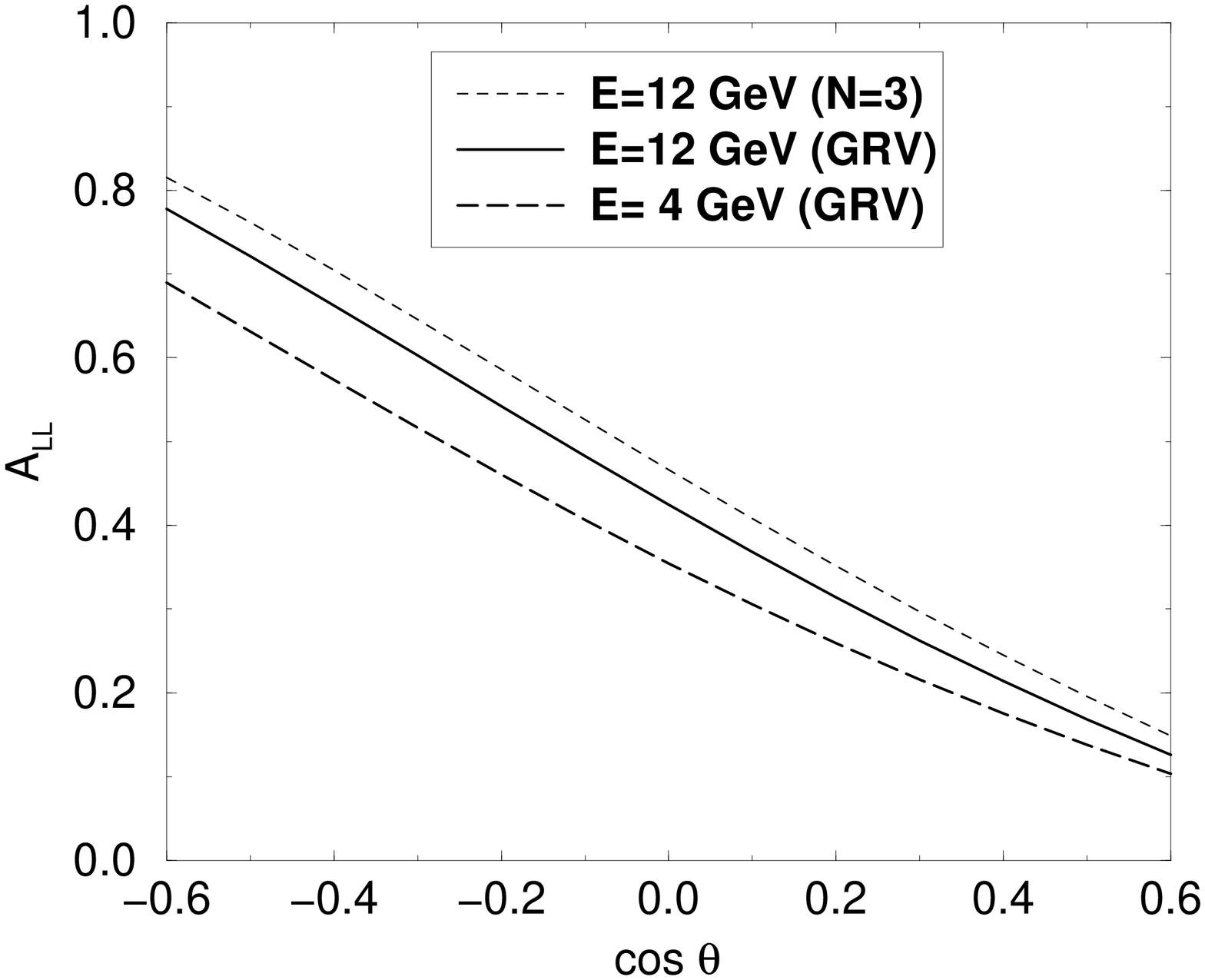, 
          width=5.7cm, angle=-90} 
\vspace*{-1cm}
\end{center}}
\caption{The Compton cross section, scaled by $s^{_6}$, (left) 
    and the initial state helicity correlation $A_{\rm LL}$ (right) 
    as predicted by the soft physics approach \cite{DFJK}. Data taken
    from \cite{shu79}. } 
\vspace*{-0.5cm}
\label{fig:all}
\end{figure}
The \spa{} also predicts characteristic spin dependences of the
Compton process. Of particular interest is the initial state
helicity correlation
\be
A_{\rm LL}\, \frac{{\d} \sigma}{{\d} t} = \frac{2\pi\aem^2}{s^2} \;
  R_V(t) R_A(t) \left(\frac{u}{s} - \frac{s}{u}\right) \,.
\ee
Approximately, $A_{\rm LL}$ is given by the corresponding subprocess
helicity correlation $\hat{A}_{\rm LL}=(s^2-u^2)/(s^2+u^2)$ multiplied
by the dilution factor $R_A(t)/R_V(t)$. Thus, measurements of both the
cross section and the initial state helicity correlation allows one to
isolate the two form factors $R_V$ and $R_A$ experimentally.
In Fig.\ \ref{fig:all} predictions for $A_{\rm LL}$ are
shown. 

The VCS cross sections have also been calculated in
\cite{DFJK}. Characteristic differences to the only other available
results, namely those from the diquark model \cite{kro91}, are to be
noticed. Thus, for instance, the beam asymmetry for $ep\to ep\gamma$
which is sensitive to the imaginary part of the
longitudinal-transverse interference, is zero in the \spa{} since all
amplitudes are real. In the diquark model, on the other hand, this
asymmetry is non-zero due to perturbatively generated phases of the
VCS amplitudes. In regions of strong interference between the Compton
and the Bethe-Heitler amplitudes the beam asymmetry is even spectacularly enhanced.

\section{SUMMARY}
The SPDs, generalized PDFs, are new tools for the description of
soft hadron matrix elements. They are central elements which connect
many different inclusive and exclusive processes: polarized and
unpolarized PDFs are the $\zeta=t=0$ limits of SPDs, electromagnetic
and Compton form factors represent moments of the SPDs, DVCS and hard
meson electroproduction are controlled by them. 
Not much is phenomenologically known on the SPDs as yet, only a few
model predictions and simple parameterizations are available. The
analysis of future data from HERA, TJlab and CERN
experiments will tell us more about the SPDs. 

A particular interesting aspect is touched in exclusive reactions such
as nucleon form factors and RCS. Their analysis by means of
SPDs implies the calculation of soft physics contributions to these
processes in which only one of the quarks is considered  as active while the others
merely act as spectators. The soft contributions formally represent
power corrections to the asymptotically leading perturbative
contributions in which all quarks participate in the elementary
scattering. It seems that for momentum transfers around 10 \gev$^2$
the soft contribution dominates over the perturbative one. However, a
severe confrontation of this approach with accurate
large momentum transfer RCS and VCS data is pending.


\begin{thebibliography}{9}

\bibitem{mue98} D.\ M\"uller, D.\ Robaschik, B.\ Geyer, F.-M.\ Dittes
                and J.\ Ho{\v{r}}ej{\v{s}}i,
                Fortschr.\ Physik {\bf 42}, 101 (1994),
                hep-ph/9812448
\bibitem{ji97}  X.\ Ji,
                Phys.\ Rev.\ Lett.\ {\bf 78}, 610 (1997);
                Phys.\ Rev.\ {\bf D55}, 7114 (1997)
\bibitem{rad97} A.V.\ Radyushkin, 
                Phys.\ Rev.\ {\bf D56}, 5524 (1997)
\bibitem{fel99} T.~Feldmann and P.~Kroll,
                hep-ph/9905343
\bibitem{ji98}  X.\ Ji and J.\ Osborne, Phys.\ Rev.\ {\bf D58}, 094018
                (1998); 
                J.C.\ Collins and A.\ Freund, Phys.\ Rev.\ {\bf D59},
                074009 (1999)
\bibitem{CFS}    A.V.\ Radyushkin, Phys.\ Lett.\ {\bf B385}, 333 (1996);   
               J.C.\ Collins, L.\ Frankfurt and M.\ Strikman, Phys.\
                Rev.\ {\bf D56}, 2982 (1997)
\bibitem{pet98} V.Yu.\ Petrov {\it et al.}, 
                 Phys.\ Rev.\ D57, 4325 (1998)
\bibitem{DFJK} M.\ Diehl, T.\ Feldmann, R.\ Jakob and P.\ Kroll,
               Eur.\ Phys.\ J. {\bf C8}, 409 (1999) and 
                hep-ph/9903268, to be published in Phys.\ Lett.\ {\bf B}
\bibitem{van99} M.~Vanderhaeghen, P.A.~Guichon and M.\ Gui\-dal,
               hep-ph/9905372
\bibitem{man98} L.~Mankiewicz, G.~Piller and T.~Weigl,
                Eur. Phys. J. {\bf C5}, 119 (1998) 
\bibitem{rad96} A.V.~Radyushkin,
                Phys.\ Lett.\ {\bf B385}, 333 (1996);
                A.D.~Martin and M.G.~Ryskin,
                Phys.\ Rev.\ {\bf D57}, 6692 (1998)
\bibitem{GRV}   M.~Gl{\"u}ck, E.~Reya and A.~Vogt,
                Z.\ Phys. {\bf C67}, 433 (1995);
                Eur.\ Phys.\ J.\ {\bf C5}, 461 (1998); 
                M.~Gl{\"u}ck, E.~Reya, M.\ Stratmann and W.\
                Vogelsang, 
                Phys.\ Rev. {\bf D53}, 4775 (1996)
\bibitem{MRST}  A.D.\ Martin {\it et al.},
                Eur.\ Phys.\ J.\ {\bf C4}, 463 (1998)  
\bibitem{CTEQ} H.L.~Lai {\it et al.}, CTEQ Collaboration,
               hep-ph/9903282
\bibitem{goe96} M.\ G\"{o}ckeler {\it et al.},
                hep-lat/9601007
\bibitem{dia96} D.I.\ Diakonov {\it et al.},
                Phys.\ Lett.\ {\bf 480}, 341 (1996)
\bibitem{sco98} S.~Scopetta, V.~Vento and M.~Traini,
                Phys.\ Lett.\ {\bf B442}, 28 (1998)
\bibitem{abo98} B.\ Abbott {\it et al.}, D0 collaboration, 
                Phys.\ Rev.\ Lett.\ {\bf 82}, 2456 (1999)
\bibitem{cdf}   J.\ Huston, CDF Collaboration, hep-ph/9901352
\bibitem{tom97} A.\ Thomas, hep-ph/9712404; U.K. Yang and A.\ Bodek,
                Phys.\ Rev.\ Lett.\ {\bf 82}, 2467 (1999)
\bibitem{sim98} G.~Ricco, S.~Simula and M.~Battaglieri,
                 hep-ph/9901360
\bibitem{kro96} R.\ Jakob and P.\ Kroll,
                Phys.\ Lett.\ {\bf B315}, 463 (1993), Erratum \ibid{} 
                {\bf B319}, 545 (1993);
                P.~Kroll and M.~Raulfs,
                Phys.\ Lett.\ {\bf B387}, 848 (1996);
                V.\ Braun and I.\ Halperin, Phgys.\ Lett.\ {\bf B328},
                457 (1994);  
                L.S.\ Kisslinger and S.W.\ Wang, Nucl.\ Phys.\ {\bf
                B399}, 63 (1993)
\bibitem{bro80}  G.P.~Lepage and S.J.~Brodsky,
                Phys.\ Rev.\ {\bf D22}, 2157 (1980)
\bibitem{CZ}    V.L.\ Chernyak and I.R.\ Zhitnitsky,
                Nucl.\ Phys.\ {\bf B246}, 52 (1984)
\bibitem{niz91} A.\ Kronfeld and B.\ Ni\v{z}i\'{c},
                Phys.\ Rev.\ {\bf D44}, 3445 (1991); 
                Erratum  {\bf D46}, 2272 (1992)
\bibitem{van97} M.\ Vanderhaeghen {\it et al.},
                Nucl.\ Phys.\ {\bf A622}, 144c (1997)
\bibitem{isg}  N.\ Isgur and  C.H.\ Llewellyn Smith, 
               Nucl.\ Phys.\ {\bf B317}, 526 (1989);
               A.V.~Radyushkin,
               Nucl.\ Phys.\ {\bf A532}, 141c (1991)
\bibitem{bol96} J.~Bolz and P.~Kroll,
                Z.\ Phys.\ {\bf A356}, 327 (1996)
\bibitem{bot89} J.\ Botts and G.\ Sterman,
                Nucl.\ Phys.\ {\bf B325}, 62 (1989);
                H.-N.\ Li and G.\ Sterman,
                Nucl.\ Phys.\ {\bf B381}, 129 (1992) 
\bibitem{ber95} J.~Bolz, R.~Jakob, P.~Kroll, M.~Bergmann and N.G.~Stefanis,
                Z.\ Phys.\ {\bf C66}, 267 (1995)
\bibitem{jai99} B.~Kundu, H.~Li, J.~Samuelsson and P.~Jain,
                Eur.\ Phys.\ J.\ {\bf C8}, 637 (1999)
\bibitem{sil93} A.F.\ Sill {\it et al.}, 
                Phys.\ Rev.\ {\bf D48}, 29 (1993)
\bibitem{rad98a} A.V.~Radyushkin,
                Phys.\ Rev.\ {\bf D58}, 114008 (1998)
\bibitem{shu79} M.A.\ Shupe {\it et al.}, Phys.\ Rev.\ {\bf D19}, 1921 (1979)
\bibitem{kro91} P.\ Kroll, M.\ Sch\"urmann and W.\ Schweiger,
                Intern.\ J.\ Mod.\ Phys.\ {\bf A6}, 4107 (1991);
                P.\ Kroll, M.\ Sch\"urmann and P.A.M.\ Guichon,
                Nucl.\ Phys.\ {\bf A598}, 435 (1996)
\end{thebibliography}
\end{document}